\documentstyle[preprint,aps]{revtex}

\newcommand{\bmq}{{\mbox{\boldmath $q$}}}

\begin{document}
\preprint {WIS-99/13 Mar-DPP}
\draft
\date{\today}
\title{R-ratios and moments of nuclear structure functions}
\author{A.S. Rinat and M.F. Taragin}
\address{Department of Particle Physics, Weizmann Institute of Science,
         Rehovot 76110, Israel}
\maketitle
\begin{abstract}

We  study  implications of  a  model,  which  links nuclear  and  nucleon
structure     functions. For it, computed     Callen-Gross      functions
$\kappa^A(x,Q^2)=2xF^A_1(x,Q^2)/F^A_2(x,Q^2)$ are for  finite $Q^2$
close to their asymptotic value
1.  Using those  $\kappa$, we compute
$R$ ratios  for $Q^2\gtrsim 5$  GeV$^2$. We review approximate methods
for the extraction of $R$ from inclusive scattering and EMC data.
We also calculate ratios of  moments of  $F^A_k$ and  find these  to
describe  the data and in particular their $Q^2$  dependence. The above
observables, as well as inclusive cross sections, are sensitive tests for
the  underlying   relation  between   nucleonic  and   nuclear  structure
functions.   In view  of the  overall  agreement, we  speculate that  the
above relation effectively circumvents a QCD calculation.

\end{abstract}

In the following we discuss two topics related to nuclear structure
functions (SF), namely ratios $R^A$ of cross sections for longitudinal
and transverse virtual photons, and ratios of moments of SF.
We start with
the cross section per nucleon for inclusive scattering of
high-energy electrons from nuclei
\begin{eqnarray}
\frac{d^2\sigma_{eA}(E;\theta,\nu)/A}{d\Omega\,d\nu}
=\frac{2}{A}
\sigma_M(E;\theta,\nu)\bigg\lbrack\frac {xM^2}{Q^2}F_2^A(x,Q^2)+
{\rm tan}^2(\theta/2)F_1^A(x,Q^2)\bigg\rbrack
\label{a1}
\end{eqnarray}
The inclusive  and the Mott  cross section $\sigma_M$  for point-nucleons
are measured as  functions of beam energy $E$,  scattering angle $\theta$
and energy loss $\nu$.  The  above nuclear SF $F_k^A(x,Q^2)$ describe the
scattering  of  unpolarized  electrons from  randomly  oriented  targets.
These depend on  the square of the 4-momentum  $Q^2=\bmq^2-\nu^2$ and the
Bjorken variable  $x$, corresponding to  the nucleon mass $M$  with range
$0\le Q^2/2M\nu\le A$.

The interest  in $F^A_k$ stems  from the interplay between  nucleonic and
sub-nucleonic dynamics which one wishes to study.  These are in principle
obtained by the Rosenbluth extraction for a single-photon exchange cross
section  (\ref{a1}), which  requires  data  for fixed  $x$  and $Q^2$  at
different      scattering     angles      $\theta$.      Since      ${\rm
sin}^2(\theta/2)=Q^2/[4E(E-Q^2/2Mx)]$,   varying  the   scattering  angle
amounts to varying the beam energy $E$.  Instead of the SF in (\ref{a1}),
one extracts the above mentioned ratio $R^A$ \cite{arneo}
\begin{mathletters}
\begin{eqnarray}
\label{a2}
R^A=d^2\sigma_L/d^2\sigma_T&=&\bigg (1+\frac{4M^2x^2}{Q^2}\bigg )\frac{1}
{\kappa^A(x,Q^2)}-1
\label{a2a}\\
\kappa^A(x,Q^2)&=&\frac{2xF^A_1(x,Q^2)}{F^A_2(x,Q^2)}
\label{a2b}
\end{eqnarray}
\end{mathletters}
We shall name $\kappa^A(x,Q^2)$ the nuclear Callen-Gross (CG) function.

There exists  a rather  extensive body  of data from  which $R$  has been
extracted, but the information does not  cover wide $x,Q^2$ ranges and is
not  accurate, reflecting  a similar  uncertainty in  $F_k^A$.  Below  we
shall discuss computed results for $R$ and standard approximations.

Eq.  (\ref{a1})  holds  irrespective   of  the  dynamics  underlying  the
description of the nuclei.  With nucleons as dominant degrees of freedom,
it is appealing  to relate SF of  nuclei to those of  nucleons, which are
considered to  be composite for  the high  $Q^2$ involved.  We  shall use
below a proposed relation \cite{gr}
\begin{eqnarray}
F_k^A(x,Q^2)&=&\int_x^{A} \frac{dz}{z^{2-k}}f^{PN}(z,Q^2)
F_k^{\langle N \rangle}\bigg(\frac{x}{z},Q^2\bigg ),
\label{a3}
\end{eqnarray}
where $F_k^{\langle N\rangle}$ are properly $p,n$-weighted SF's of $free$
nucleons $F_k^p,F_k^n\approx  F_k^D/2-F_k^p$.  Those  contain information
on the  sub-structure of the nucleon  and we shall use  data compiled for
$F_1^N$, and  parametrizations for  $F_2^N$ \cite{bod}.   Dynamics enter
through the SF of a nucleus with point-particles $f^{PN}$, probed at high
$Q^2$.   The  relation  (\ref{a3})  is  thought  to  be  valid  for  both
$Q^2\gtrsim$(1-1.5)GeV$^2$ and for $x\gtrsim 0.15$, below which neglected
pionic \cite{pion}  and anti-screening effects grow in  importance.  In
addition, $A\gtrsim  12$  in  view  of the  neglect  of  nucleon  recoil.
Applications to cross sections data \cite{na3,arr} have met with definite
success \cite{rt,rt1}.

We  have  demonstrated   before  that the above $f^{PN}$ is  only  weakly
$A$-dependent as are the  weighted $F_k^{\langle N\rangle}$, even for
the  largest  neutron  excess   $\delta  N/A$.   Eq.  (\ref{a3})  through
(\ref{a2}) then implies
\begin{eqnarray}
\kappa^A=\kappa^{\langle N\rangle} +{\cal O}(1/A)&\approx&
\kappa^D(x,Q^2)+{\cal O}(1/A)
\nonumber\\
R^A(x,Q^2)&\approx& R(x,Q^2)+{\cal O}(1/A),
\label{a4}
\end{eqnarray}
in agreement with data \cite{arneo,das1}. Using first
the CG $relation$  for nucleons
\begin{eqnarray}
\epsilon^N_{CG}=\lim_{Q^2\to \infty} \kappa^N(x,Q^2)=1
\label{a5}
\end{eqnarray}
one finds from (\ref{a2b}) and (\ref{a4}), its nuclear analog,
\begin{eqnarray}
\epsilon^A_{CG}=\lim_{Q^2\to \infty} \kappa^A(x,Q^2)=1+{\cal O}(1/A)
\label{a6}
\end{eqnarray}
With  (\ref{a5}), the  nuclear  CG $relation$  (\ref{a6})  can be  proven
directly from (\ref{a3}).  In  contradistinction, the equality of nuclear
and nucleonic  CG $functions$  (\ref{a4}) is compatible  with (\ref{a3}),
but does not necessarily follow from it.

First we mention a remarkable observation for the computed CG functions
\begin{eqnarray}
|\kappa(x,Q^2)-1|&\approx&(0.11-0.12)
\nonumber\\
(0.2-0.3) \lesssim x& \lesssim&(0.7-0.75);\,\,Q^2\ge 5 {\rm GeV}^2
\label{a7}
\end{eqnarray}
In the indicated  $x$-interval and over a wide  $Q^2$-range, CG functions
appear to  be close to their  asymptotic limit, the nuclear  CG relation.
It is  also intriguing  that without  any apparent  cause, a  sign change
occurs at a weakly $Q^2$-dependent $x_s\approx 0.5-0.6$.  The above is in
agreement with  data from high energy  $\nu,\bar\nu$ inclusive scattering
(see Fig.  18 in  \onlinecite{berg}).  The  small $  \kappa-1 $  shall be
shown to entail disproportionally large effects.  For later use we remark
on the estimated accuracy of  the computed CG function (\ref{a2b}), which
appears limited in various ranges:

i) Disregarding other than valence quarks, requires smoothing of $F_k^N$
for $x\lesssim$ 0.15-0.20,  which entails the same for  $F_k^A$.  We thus
prefer  to  use  extrapolated  values for  nuclear  CG  functions,  below
$x\lesssim$0.15.

ii) Eq. (\ref{a3})  shows that $f^{PN}$ draws on an  ever smaller support
of dwindling intensity and  accuracy, rendering $F_k^A(x,Q^2)$ unreliable
beyond $x\ge 1.3-1.5$.

iii) The parametrizations for $F_2^p\,,F_2^D$ \cite{bod} hold for $Q^2\le
$20 GeV$^2$, causing uncertainties in $F_k^A$ for larger $Q^2$.

iv)  With SF  for  $x\ge 1.2$  falling orders  of  magnitudes from  the
maximum values,
one  expects inaccuracies  if $F_k^A$  and $\kappa$  for growing
$x$.

We now discuss three approximations  $R_n$ for $R^A\approx R$, defined by
a corresponding choice  for the CG function  $\kappa_n$.  For each of
these one has from (\ref{a2})
\begin{eqnarray}
R(x,Q^2)=\beta_n(x,Q^2)R_n(x,Q^2)+\bigg
(\beta_n(x,Q^2)-1   \bigg)
\label{a8}
\end{eqnarray}
Deviations of $\beta_n(x,Q^2)=\kappa_n(x,Q^2)/\kappa(x,Q^2)$ from
1 manifestly determine the quality of the approximation.

A) A high-$Q^2$ approximation, defined by $\kappa_L=1$ (i.e. $\beta_L=
\kappa^{-1}$), approximately valid for $1 \lesssim x\lesssim $0.6:
\begin{mathletters}
\label{a9}
\begin{eqnarray}
R^{'exact'}(x,Q^2)&=&
\beta_L(x,Q^2)R_L(x,Q^2)+\bigg (\beta_L(x,Q^2)-1 \bigg)
\label{a9a}\\
&\approx& R_L(x,Q^2)+\bigg (\beta_L(x,Q^2)-1\bigg )
\label{a9b}\\
R_L^{(1)}(x,Q^2)&=&\frac{4M^2x^2}{Q^2}+ (\beta_L(x,Q^2)-1)
\label{a9c}\\
R_L^{(2)}(x,Q^2)&=&\frac{4M^2x^2}{Q^2}
\label{a9d}
\end{eqnarray}
\end{mathletters}
Eq. (\ref{a9a}) is the same as Eqs.  (\ref{a2}).  The corresponding
$R$ is  dubbed $'$exact$'$,  because it results  from computed  values of
$F_k^A$, Eq. (\ref{a3}) \cite{rt1}, which implies some model. $R^{'{\rm
exact}'}$ should be distinguished from intrinsic approximations for $R$.

B) The  NE approximation  for $x\approx1$ rests  on the  decomposition of
$F_k^N$  in  (\ref{a3})  into  $p,n$-weighted  nucleon-elastic  (NE)  and
nucleon-inelastic (NI) parts.  Retention of the NE part generates through
(\ref{a3})  corresponding  NE  parts  in  the  $nuclear$  SF,  thus  with
$\eta=Q^2/4M^2$
\begin{mathletters}
\label{a10}
\begin{eqnarray}
F_1^{N(NE)}(x,Q^2)&=&\frac{x}{2}[G^N_M(Q^2)^2\delta(x-1)
\nonumber\\
F_2^{N(NE)}(x,Q^2)&=&\frac{[G^N_E(Q^2)]^2+\eta[G_M^N(Q^2)]^2}{1+\eta}
\delta(x-1)
\label{a10a}\\
F_1^{A(NE)}(x,Q^2)&=&\frac{1}{2}f^{PN}(x,Q^2)[G^N_M(Q^2)]^2
\nonumber\\
F_2^{A(NE)}(x,Q^2)&=&xf^{PN}(x,Q^2)
\frac{[G^N_E(Q^2)]^2+\eta[G_M^N(Q^2)]^2}{1+\eta}
\label{a10b}
\end{eqnarray}
\end{mathletters}
The corresponding CG function can be simplified by exploiting the
approximate scaling of the static electro-magnetic form factors in the
NE part (\ref{a10a}), $1/[(\mu_M^p)^2+(\mu_M^n)^2]=0.0874$ \cite{bos}
\begin{eqnarray}
\kappa^A_{NE}&=&2xF_1^{A(NE)}/F_2^{A(NE)}
\nonumber\\
&\approx&(0.0874+\eta)/(1+\eta)
\label{a11}
\end{eqnarray}
Inserting (\ref{a11}) into (\ref{a3}) gives
\begin{mathletters}
\label{a12}
\begin{eqnarray}
R(x,Q^2)&=&\beta_{NE}(x,Q^2)R_{NE}(x,Q^2)
+\bigg (\beta_{NE}(x,Q^2)-1 \bigg)
\label{a12a}\\
R_{NE}^{(1)}(x,Q^2)
&=& \frac{0.31}{Q^2}+\bigg (\frac{0.31}{Q^2}+1\bigg )
\bigg ( \frac{x^2-1}{1+\eta}\bigg ),
\label{a12b}\\
R^{(2)}_{NE}(x,Q^2)&\approx& \frac{0.31}{Q^2},
\label{a12c}
\end{eqnarray}
\end{mathletters}
with $Q^2$ expressed in GeV$^2$.
Eq. (\ref{a12c}) is the result of Bosted et al \cite{bos}, while
Eq. (\ref{a12b}) provides $x$-dependent corrections.

C) An empirical estimate for moderate $Q^2$, which is assumed
to be independent of $x$ and $A$ \cite{bos,das2,brad}
\begin{eqnarray}
R_C(x,Q^2)\approx\frac{\delta}{Q^2}\,\,\,\,\,; 0.2 \lesssim \delta
\lesssim 0.5,
\label{a13}
\end{eqnarray}
The estimates (\ref{a9d}), (\ref{a12c}) for $x\approx 1$, and (\ref{a13})
predict  $R\propto 1/Q^2$,  but only  A) and  B) for  $x\ne 1$  prescribe
definite $x$ dependence.   Since by definition $R$ depends on  $x$, it is
likely that extracted coefficients of $1/Q^2$ effectively hide
actual $x$-dependence.

Were it  not for the listed inaccuracies  in computed CG functions,
the latter would through (\ref{a2a}) or (\ref{a9a}) provide a
standard for all approximate $R$ ratios.   We now discuss those and start
with  the  large  $Q^2$  approximation.    In  view  of  the  observation
(\ref{a7}), the  CG function  $\kappa\approx 1$  holds also  for moderate
$Q^2$   and    over   a   relatively   wide    $x$-range.    For   $'{\rm
medium}'\,\,x^2/Q^2$,  which does  not require  large $Q^2,\,\,\,R\approx
R^{(2)}$ may  suffice.  However, in  the deep-inelastic region  for small
enough $x^2/Q^2$, even  for a few \% deviation of  $\beta_L$ from 1, the
second part  in (\ref{a9c})  exceeds $R_L^{(2)}$, and  (\ref{a9c}) should
therefore be used there.

In Table I we present  results for relatively low $x$,
$0.12\lesssim x\lesssim 0.7$  and for
$Q^2\ge  5$ GeV$^2$.   The first  row gives  $R^{'exact'}$, Eq.
(\ref{a9a}),  computed from  (\ref{a3}),  except the  entry for  $x=0.12$
which,  as explained  above,  has been  extrapolated  down from  slightly
larger  $x$.  The  second row  is the  asymptotic limit  $R_L^{(2)}$, Eq.
(\ref{a9d}).  We do not display $R_L^{(1)}$, since it virtually coincides
with  $R^{'exact'}$.  One  notices that  for higher  $x$, the  asymptotic
limit is either close to, or  exceeds the exact answer.  This reflects on
$\kappa$,  Eq. (\ref{a2b})  to  be  close to,  or  exceeding  1, in  turn
entailing a  negative correction  to $R_L^{(2)}$.   This agrees  with the
observation  (\ref{a7}).   The  last  column  contains  a  few  scattered
$\nu,\bar\nu$ data for the indicated  $x$ and binned $\langle Q\rangle^2$
\cite{benv,berg}.   Given the  substantial statistical  and systematic
errors and the imprecisely given
spreading due to binning, the agreement is reasonable.

Next  we discuss  the NE  approximation,  the validity  of which  depends
foremost on  the weight of  $F_k^{N(NE)}(x,Q^2)$ in $F_k^A$.   When using
(\ref{a3}), that  weight is  determined by $f^{PN}$,  for which  there is
only   theoretical  information.    Computations  show   that  only   for
$Q^2\lesssim  2$ GeV$^2$, $F_k^{A(NE)}(x,Q^2)$ dominates for
$ x \lesssim $(1.1-1.2). For  growing $Q^2$ NI parts  compete for ever
growing  $x$ and  ultimately overtake \cite{rt1}.

Disregarding NI contributions to  $R_{NE}$ for $x\ne 1$, corrections
in the  immediate neighborhood of  the QEP  can be estimated  by choosing
$\beta_{NE}$ close  to 1.   One thus finds  $R(1.05,5)/R^{NE}(1,5)$= 1.86
which ratio  rapidly increases with  $\beta_{NE}$.  One also  checks from
(\ref{a12b})   that   for   $1\lesssim   Q^2({\rm   GeV}^2)\lesssim   5$,
$R_{NE}(x\lesssim 0.9,Q^2)$ reaches unphysical negative values.  Only the
disregarded NI part  can restore $R$ to positive values.
For $1.5  \gtrsim Q^2({\rm GeV}^2)\gtrsim 5$ and for instance $x=$1.1 on
the elastic side  of the  QEP,
$2 \gtrsim R_{NE}(1.1,Q^2)/R_{NE}(1,Q^2)\gtrsim 1.5$,
which ratio again grows  with $x$: NI terms may, or  may not off-set that
growth.      Table     II      compares     the     NE     approximations
$R^{(1)}_{NE},R_{NE}^{(2)}$  with  $R_C$:  the  agreement  is  tolerable.
Aware  of the  warnings after  (\ref{a7}), we  nevertheless compute  and
enter some $'$exact$'$ values, which appear  to exceed the NE values by
far.  CG  functions $\kappa(1,Q^2)$ which  fit $R_{NE}$ would have  to be
25-30 \% larger  than the computed ones, which we  estimate to be outside
the limits of our accuracy.  In particular the negative $R_{NE}(0.9,Q^2)$
makes one believe that the NE estimates may not be precise.

Eq.  (\ref{a12c})  has been  applied  to  extract  $R$ and  $F_2^A$  from
inclusive scattering data  for medium-$Q^2$ data for  $x\approx 1$
\cite{brad,arr1}. Data
by Bosted  et al for  0.75$\lesssim x\lesssim$1.15 are quite  eratic, but
$R(\langle x\rangle,Q^2)$, averaged over $x$,  shows a trend in agreement
with (\ref{a12c}).

In  addition there  are data  for about  the same  $Q^2$-range, but  more
restricted  $x$   \cite{arr1},  which   are  in  agreement   with  either
(\ref{a12c}) or ({\ref{a13}).  There  clearly are substantial corrections
just off the QEP.  In particular for
the data of Bosted et al, the above
warns  that the  use of  simple $x$-independent  $R$ ratios  may lead  to
extracted  $F^A_2$, which  have inaccuracies, exceeding those estimated.

\bigskip
\bigskip

We now address a second topic  regarding the moments of various SF
\begin{eqnarray}
{\cal M}_k^A(m;Q^2)&=&\int_0^A dx x^m F_k^A(x,Q^2)
\nonumber\\
{\cal M}_k^N(m;Q^2)&=&\int_0^1 dx x^m F_k^N(x,Q^2)
\nonumber\\
\mu^A(m;Q^2)&=&\int_0^A dx x^m f^{PN}(x,Q^2)
\label{a14}
\end{eqnarray}
Moments ${\cal M}_k^N$  describe higher twist corrections of  SF of
nucleons \cite{pen}, and the same holds for their nuclear
counterparts, had those been calculated in QCD. Our interest in those
moments is  the sensitivity of  SF for large $x$ and consequently the
trust in the calculated $F^A_k$ for that range.
One readily derives from (\ref{a10}) \cite{foot2}
\begin{mathletters}
\label{a15}
\begin{eqnarray}
F_k^A(0,Q^2)&=&\mu^A(-2+k;Q^2)F_k^N(0,Q^2)
\label{14a}\\
{\cal M}_k^A(m,Q^2)&=&\mu^A(m-1+k;Q^2){\cal M}_k^N(m;Q^2)
\label{14b}\\
\mu^A(m+1;Q^2)&=&\frac {{\cal M}_1^A(m+1;Q^2)}{{\cal M}_1^N(m+1;Q^2)}
=\frac {{\cal M}_2^A(m;Q^2)}{{\cal M}_2^N(m;Q^2)}
\label{a15c}
\end{eqnarray}
\end{mathletters}
and in particular
\begin{eqnarray}
\mu^A(0,Q^2)=\int_0^A dx f^{PN}(x,Q^2)=\int_0^A dx f^{as}(x)=1,
\label{a16}
\end{eqnarray}
which expresses  unitarity.  All  other relations (\ref{a15})  for finite
$Q^2$ rest  on the  representation (\ref{a3}) and  embody effects  of the
binding medium on moments of $F_k^N$ through $\mu(n,Q^2)$.  For instance,
the deviation  of $\mu^A(2,Q^2)$  from 1 measures  the difference  of the
momentum fraction  of a quark  in a nucleus and  in the nucleon  at given
$Q^2$.

We  have computed  the  lowest  moments and  ratios  $\mu$ from  computed
$F_k^A, f^{PN}$ and parametrized  $F_k^N$.  With expected inaccuracies in
$F_k^A$  for $x\gtrsim  1.5$ one  ought  not to  trust calculated  higher
moments.   Yet we  found consistent  values for  the different  ratios in
(\ref{a15c}) for $Q^\le 20$ GeV$^2$, and the moments of $f^{PN}$.
Those for  Fe are entered  in Fig. 1 and  agree reasonably well  with the
available  data.  We  note in  particular the  rendition of  the observed
$Q^2$-dependence, as opposed to a  similar investigation by Cothran et al
\cite{coth}.  The authors used a generalized convolution like (\ref{a3}),
with  a $Q^2$-independent  PWIA for  $f^{PN}$,  leading to  the same  for
$\mu(m)$. $Q^2$-dependence, estimated for off-shell nucleons, produce far
too small moment ratios with the wrong $Q^2$ behavior.

The above is reminiscent of
previously considered, but not identical moments. We recall
discrepancies between data and computed results for relatively
low-$q$, longitudinal
responses  $S_L$ and the integral  of the latter, the  Coulomb sumrule
\onlinecite{mez,coh}.   All  have  occasionally   been  ascribed  to  the
influence of  the binding medium  on the size of  a nucleon, i.e.  on the
second  moment  of the  $static$  charge  density.  Apart  from  possible
conventional accounts of those differences \cite{jour}, one notes that
(\ref{a3}) does not relate to static moments of charge distributions, but
to dynamical SF.

The above and  Refs. \onlinecite{rt,rt1} conclude a  program to determine
observables which depend  on nuclear SF, in turn computed  from the basic
relation (\ref{a2}) between SF for composite nuclei, free nucleons and of
a  nucleus   composed  of   point  nucleons.   The   various  observables
occasionally extend  over wide ranges,  and test to various  measures the
$x,Q^2$ dependence of $F^A_k$.  It  is gratifying to frequently note good
agreement with data.

The  above clearly  requires an  explanation, because  results have  been
obtained, circumventing QCD.  It seems  hard to avoid the conclusion that
in the  tested $x,Q^2$ region,  the relation  (\ref{a2}) is result  of an
effective theory,  as has  been argued  originally \cite{gr}  and somehow
mimicking notions of QCD.

\bigskip
\par

\newpage

%\end{document}

%\documentstyle[preprint,prb,aps]{revtex}
%\baselineskip 16pt
%\newpage
%\makebox[8.5in]{\hfill{\bf Table II}\hfill}
%\begin{minipage}[b]{8.5in}
\begin{center}
{\bf Table I}
\vskip 1cm
%\vspace{1cm}
\begin{tabular}{|c||c||c|c|c|c|}
\hline
             & $Q^2$(GeV$^2$)&    5    & 10    &20      &  50  \\
\hline
\hline
    $x$      &    $R$       &         &        &        &      \\
\hline
\hline
             &  $R^{'exact'}$&  0.284 & 0.226  & 0.221  &0.218 \\
   0.08   &  $R_L^{(2)}$  &   0.005 & 0.002  & 0.001  &0.000 \\
     &$R^{exp}(\langle Q^2\rangle\approx 7)$&
       \multicolumn{2}{c|}{$0.27\pm 0.06\pm 0.02$} & &
     \\
\hline
             &  $R^{'exact'}$&  0.216& 0.203  &  0.185 & 0.176\\
   0.12      &  $R_L^{(2)}$  &   0.010 & 0.005  & 0.003  &0.001 \\
     &$R^{exp}(\langle Q^2\rangle\approx 12)$& &
\multicolumn{2}{c|}{$0.12\pm 0.05\pm 0.02$}  &       \\
\hline
             &  $R^{'exact'}$&  0.192 & 0.169 & 0.146  &0.120 \\
   0.18      &  $R_L^{(2)}$ &   0.023 & 0.011  & 0.005  &0.002 \\
     &$R^{exp}(\langle Q^2\rangle)\approx 23)$& &  &
\multicolumn{2}{c|}{$0.06\pm 0.06\pm 0.02$}   \\
\hline
             &  $R^{'exact'}$&   0.159 & 0.120  & 0.089 &0.044 \\
   0.27      &  $R_L^{(2)}$ &   0.051 & 0.025  & 0.013  &0.006 \\
   &$R^{exp}(\langle Q^2\rangle\approx 30)$&  &   &
\multicolumn{2}{c|}{0.04$\pm 0.04\pm 0.01$} \\
\hline
             &  $R^{'exact'}$&   0.144 & 0.119  & 0.064  & 0.009 \\
   0.36      &  $R_L^{(2)}$ &   0.091 & 0.025  & 0.013  &0.006 \\
&$R^{exp}(\langle Q^2\rangle\approx50)$& & & & $\,\,\,\,$
                   $-0.04\pm 0.04\pm 0.01$\\
\hline
   0.5       &  $R^{'exact'}$&  0.140 & 0.113  & 0.048  &$\approx$ 0 \\
             &  $R_L^{(2)}$ &   0.178 & 0.089  & 0.044  &0.018 \\
\hline
   0.7       &  $R^{'exact'}$&  0.223 & 0.170  & 0.120  &$\approx$ 0 \\
             &  $R_L^{(2)}$ &   0.348 & 0.170  & 0.085  &0.035 \\
\hline
\end{tabular}
\end{center}
%\vspace{1cm}
\vskip 1cm
$'$Exact$'$
$R$ for low $x$ and medium-high $Q^2$, the high $Q^2$ limit
and data for binned $\langle Q^2\rangle$ \cite{benv,berg}.
The first row for  $x=0.1$  are extrapolations down to $x=0.1$.
\newpage

\begin{center}
{\bf Table II}
\vskip 1cm
%\vspace{1cm}
\begin{tabular}{|c||c|c|c|c|}
\hline
$x$  & $R$     & $Q^2$(GeV$^2$): 2   & 5    & 10     \\
\hline
\hline
   0.9       &  $R_{NE}^{(1)}$&  $<0$        &$<0$    &  $<0$   \\
             &  $R_{NE}^{(2)}$&  0.155      &0.062   &  0.032  \\
             &  $R^{'exact'}$&    -        & 0.292     &   0.308  \\
\hline
   1.0       &  $R_{NE}^{(1)}$& 0.155       & 0.062     &   0.032    \\
             &  $R_{NE}^{(2)}$& 0.155       & 0.062     &   0.032  \\
             &  $R^{'exact'}$&    -        & 0.329     &   0.404  \\
\hline
   1.05      &  $R_{NE}^{(1)}$&  0.231      & 0.117     & 0.059   \\
             &  $R_{NE}^{(2)}$&  0.155      & 0.062     & 0.031  \\
\hline
\hline
  $x$        &$R_C([0.4\le\delta \le0.6])$&0.2-0.3&0.08-0.12& 0.04-0.06\\
\hline
\end{tabular}
\end{center}
%\vspace{1cm}
\vskip 1cm
$R$ ratios (\ref{a12b}), (\ref{a12c}) for $x\approx 1$ , medium-$Q^2$
and the $x$-independent $R_c$, Eq. (\ref{a13}). For $x=0.9,1.0\,\,;
Q^2$=5,10 GeV$^2$
we also entered $R^{'exact'}(x,Q^2)$. See text for discussion.

{\bf figure captions.}

{Fig. 1}
Second, third and fourth moments $\mu(m,Q^2)$, Eq. (\ref{a16}).


\begin{references}
\bibitem{foot2}
We disregard  here the so-called
flux  factor,  holding  that  $f(z)\to  f(z)/z$,  or  equivalently,
$\mu(m+1)\to \mu(m)$.   Numerical consequences  are anyhow  minute, since
$\mu(m)$ changes  only gently with  $m$.
\end{references}
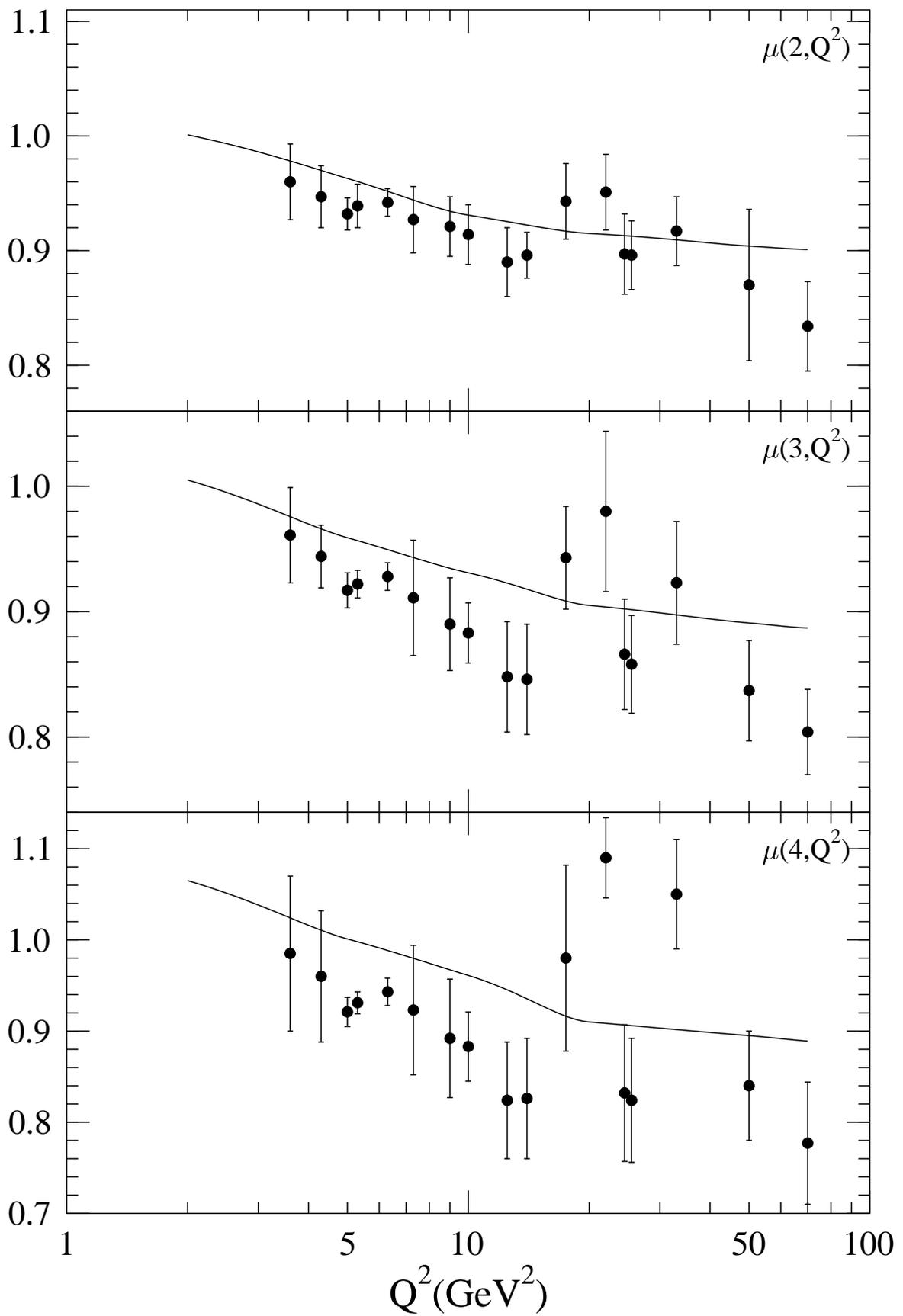
\end{document}